\documentclass{an}

\usepackage{ulem}
\usepackage{color}
\usepackage{graphics,graphicx}
\usepackage{times} 
\usepackage{amssymb}
\usepackage{lscape}
\usepackage{url}
\usepackage{multirow}
\usepackage{ctable}
\usepackage{threeparttable}
\newif\ifAMStwofonts
\AMStwofontstrue
\title{A Large Catalogue of Ultraluminous X-ray Source Candidates in Nearby Galaxies}

\author{D.J. Walton\inst{1}\fnmsep\thanks{\email{dwalton@ast.cam.ac.uk}}
\and  J.C. Gladstone\inst{2}
\and  T.P. Roberts\inst{3}
\and  A.C. Fabian\inst{1}
}
\titlerunning{A Large Catalogue of ULX Candidates in Nearby Galaxies}
\authorrunning{D.J. Walton, \textit{et al.}}
\institute{
Institute of Astronomy, Cambridge University, Madingley Road, Cambridge, CB3 0HA
\and 
Department of Physics, University of Alberta, Edmonton, Alberta, T6G 2C7, Canada
\and
Department of Physics, University of Durham, South Road, Durham, DH1 3LE}

\received{TBC}
\accepted{TBC}
\publonline{TBC}

\def\xmm{{\it XMM-Newton~\/}}

\def\nustar{{\it NuSTAR}}

\def\epicpn{{\it EPIC}{\rm-pn}}
\def\epicmos1{{\it EPIC}{\rm-MOS1~\/}}
\def\epicmos2{{\it EPIC}{\rm-MOS2 ~\/}}
\def\epicmos{{\it EPIC}{\rm-MOS}}

\def\apj{ApJ}
\def\mnras{MNRAS}

\def\aap{A\&A}                   
\def\apjs{ApJS}                  
\def\apjl{ApJ}                   
\def\pasj{PASJ}

\def\kmps{\hbox{$\rm\thinspace km~s^{-1}$}}

\def\H0{{\rm ~km~s^{-1}~Mpc^{-1}}}

\def\kev{\hbox{$\rm\thinspace keV$}}

\def\atpcm{{atom~cm$^{-2}$}}

\def\ergps{\hbox{erg~s$^{-1}$}}

\def\grid25{\hbox{\rm{\small GRID25}}}

\def\etal{et al.~\/}
\def\eg{{\it e.g.~\/}}

\def\ie{{\it i.e.~\/}}

\def\la{\mathrel{\hbox{\rlap{\hbox{\lower4pt\hbox{$\sim$}}}{\raise2pt\hbox{$<$}}}}}
\def\ga{\mathrel{\hbox{\rlap{\hbox{\lower4pt\hbox{$\sim$}}}{\raise2pt\hbox{$>$}}}}}

\def\d25{D$_{25}$}
\def\nh{{$N_{H}$}}

\def\.25{0.25 keV\thinspace}

\def\rg{$R_{G}$}

\def\ulxone{\rm{NGC\thinspace4517 ULX1}}
\def\ngc4517{\rm{NGC\thinspace4517}}


\abstract{%
  Since their discovery, Ultraluminous X-ray sources (ULXs) have attracted attention due
  to their combination of extreme luminosities and extra-nuclear locations. However, they
  are a fairly rare phenomenon, and attempts to investigate the general properties of
  the population have been hindered by a relative lack of known sources.
  Here, we present a large catalogue of ULX candidates including 655 detections of 475
  discrete sources, based on the 2XMM Serendipitous Survey. To demonstrate the potential
  of such a resource, we present some scientific analysis of this population, focusing on
  the spectral turnover seen, often at $\sim$6\,\kev, in the highest quality ULX data. We also
  demonstrate how the recent reflection and Comptonisation interpretations of this
  feature may be distinguished observationally in the future, specifically using \ulxone,
  a previously unanalysed source with high quality data, as an example case.}

\begin{document}
\maketitle
\label{firstpage}

\section{Introduction}

\begin{figure*}
\begin{center}
\rotatebox{0}{
{\includegraphics[width=230pt]{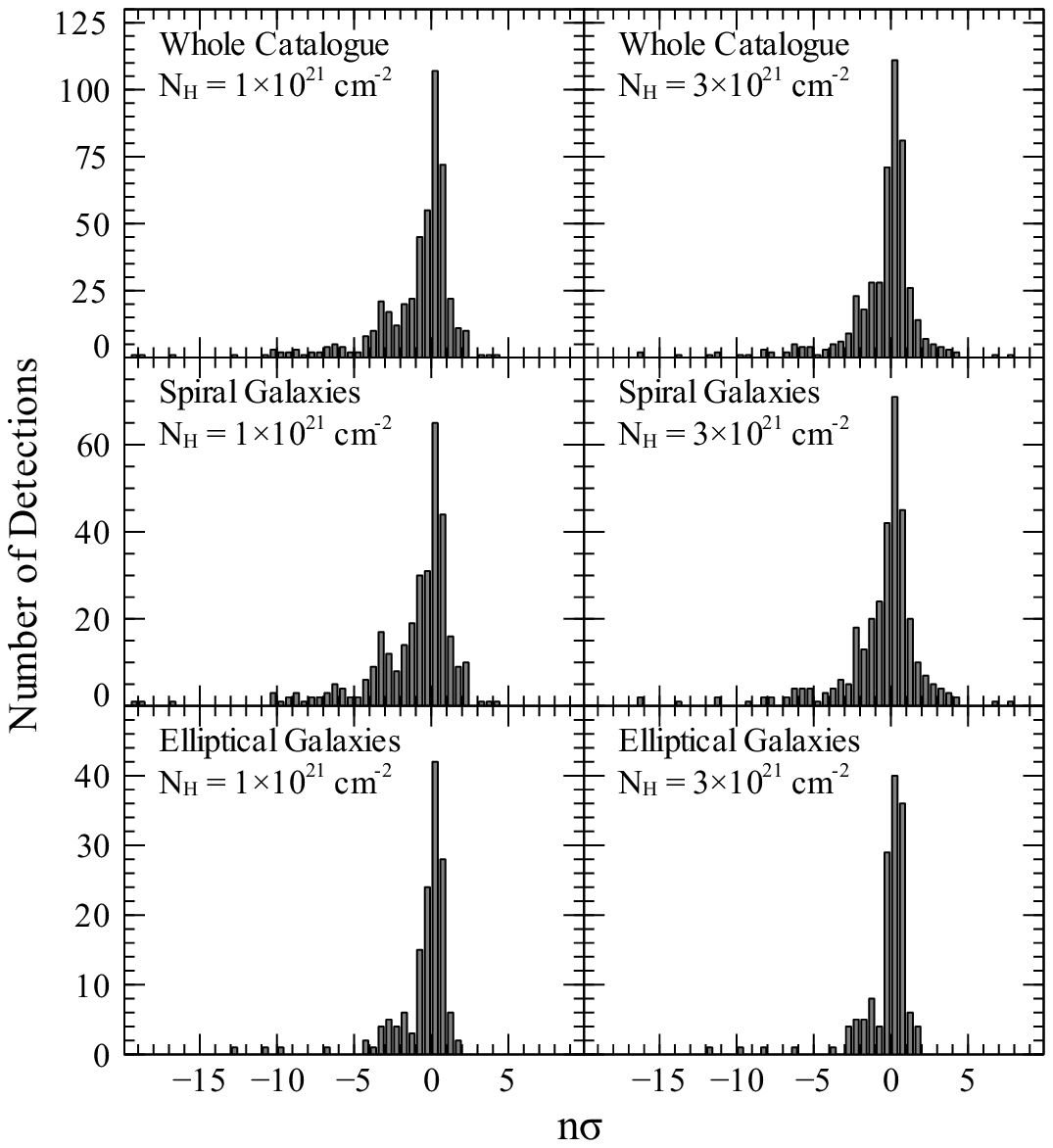}}
}
\rotatebox{0}{
{\includegraphics[width=230pt]{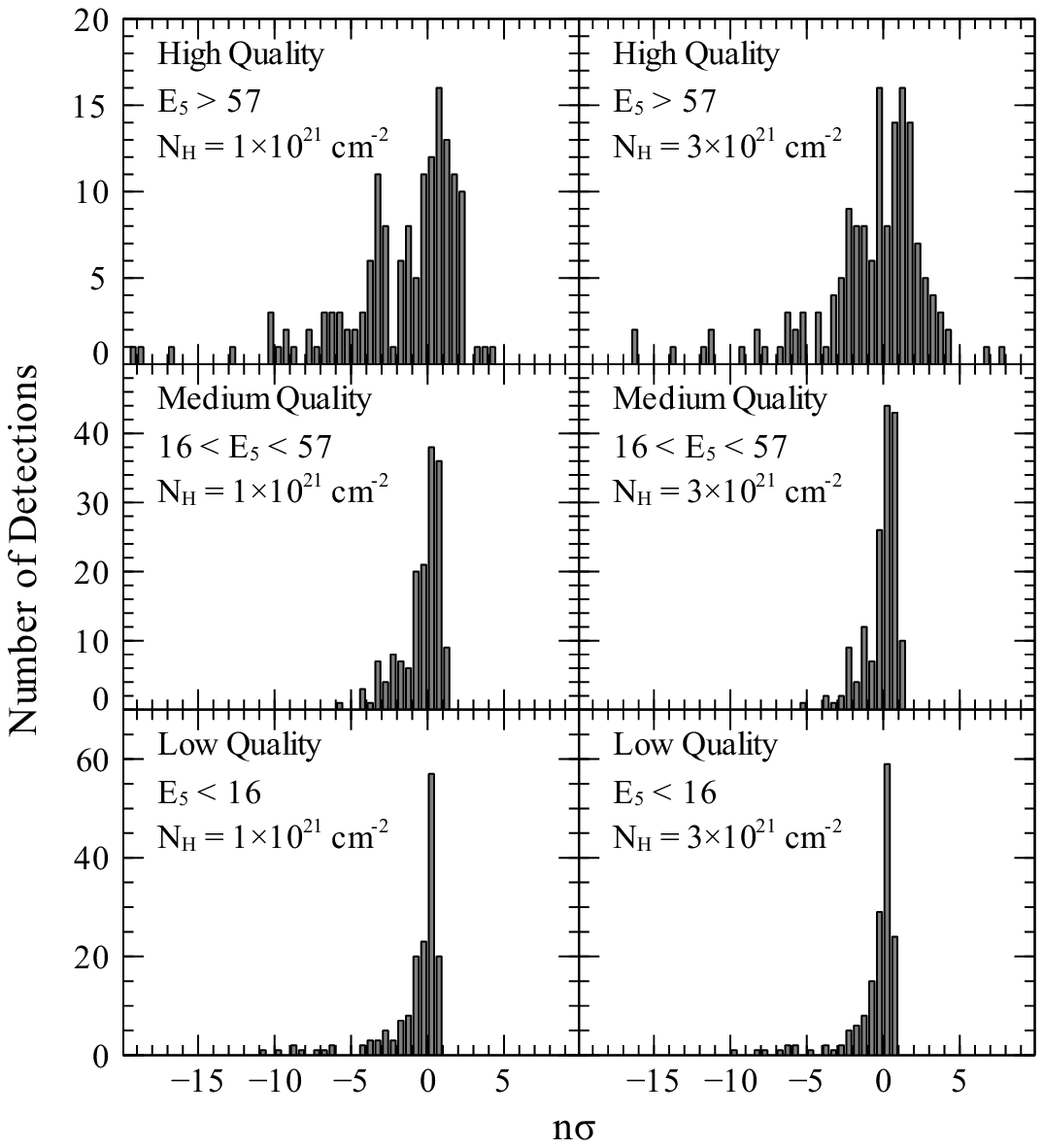}}
}
\end{center}
\caption{Number distributions for detections in the presented catalogue, and
some of its subsets, of the $n\sigma$ deviation of the observed \epicpn\
band 5 counts from the expected counts calculated by extrapolating the lower
energy spectral shape, assuming a simple absorbed power-law model. Left panel:
distributions for the whole catalogue (top row), and the spiral and elliptical
galaxy subsets (middle and bottom rows respectively). Right panel: distributions
for the high, medium and low quality datasets (top, middle and bottom rows
respectively; determined by band 5 EPIC counts, $E_5$). Both panels: distributions
in the left and right columns are calculated assuming $\langle$\nh$\rangle$ = 1
and 3 $\times10^{21}$ \atpcm. Most distributions are significantly skewed and
extended towards negative values of $n$, which suggests the presence of a
spectral turnover similar to that seen in the highest quality ULX data in a
significant fraction of our population.}
\label{fig_turn}
\end{figure*}

Ultraluminous X-ray sources (ULXs) display the unusual combination of intense X-ray
luminosities (exceeding $10^{39}$ \ergps) and extra-nuclear locations, and as such
have attracted considerable attention as attempts are made to identify the cause of
their apparently extreme emission. It is well accepted that the majority of
ULXs are likely to be accreting black holes, and most explanations put forward are
based on one (or more) of the following ideas: ULXs may contain larger black holes than
stellar-mass black holes hosted by typical X-ray binaries (XRBs), possibly intermediate-mass
black holes (Colbert \& Mushotzky 1999; dynamical friction arguments almost certainly
rule out the majority of ULXs hosting supermassive black holes; Miller \& Colbert 2004), they may be
stellar-mass black holes radiating above their Eddington limit (Begelman 2002), or
they may be sources that do not emit isotropically, hence luminosities calculated assuming
isotropy overpredict their power output (King \etal 2001). In the latter case,
the optical nebulae observed around many ULXs seem to rule out extreme anisotropy due
to relativistic jets orientated towards us (\eg a `micro-blazar'; Pakull \& Gris\'e
2008), but modest collimation by \eg `slim' accretion discs remains an intriguing
possibility.

Observationally, it has proven difficult to distinguish between these explanations,
notably between the IMBH and stellar mass interpretations. For a recent review of this
topic, see Roberts (2007). In large part, this is because reliable dynamical mass
estimates are not yet available for any ULXs. However, ULXs are a fairly rare phenomenon,
and a relative lack of known sources on which to base population studies has also hindered
progress. Here, we present a large catalogue of X-ray detections of ULX candidates,
as well as some analysis of its population, as a step towards addressing this issue.

\section{The Catalogue}
\label{sec_red}

The main resource used in this work is the 2XMM Serendipitous Survey (2XMM; Watson \etal 2009).
In the interest of brevity, here we list only the main steps taken to condense 2XMM
into a catalogue of ULX candidates. For a full description of the data reduction
see Walton \etal (in prep.).

\begin{itemize}
 \item 2XMM was cross-correlated with the Third Reference Catalogue of Bright Galaxies
(RC3; de Vaucouleurs \etal 1991), in order to select the X-ray sources within the elliptical $D_{25}$
galaxy regions.
 \item X-ray luminosities were calculated using the full band (0.2--12.0\thinspace \kev) EPIC flux
and galactic distances taken from Tully (1988) for galaxies with a recession velocity $cz < 1000$
\kmps, or calculated assuming the Hubble flow otherwise (adopting $H_{0} = 75\H0$ for consistency
between the two regimes).
 \item Sources flagged as extended, and those fainter than $10^{39}$ \ergps\ by more than
1$\sigma$, were discarded.
 \item A minimum detection significance of 3.5$\sigma$ was required, to reduce the contribution
from spurious detections.
 \item Sources within 7.5'' of the galactic centre were excluded to minimize the contribution of
low luminosity AGN.
 \item Known contaminants (\eg AGN, SNe) were removed.
\end{itemize}

These filtering measures produced a catalogue of 655 X-ray detections of 475 ULX
candidates; 59 of these sources are nominally fainter than $10^{39}$ \ergps\ but
agree with this limit within their 1$\sigma$ luminosity uncertainties. Of these 475
candidates, 307 are located in 142 spiral galaxies, with the remaining 168 located
in 98 elliptical galaxies, despite the observations contributing to 2XMM covering
twice as many ellipticals than spirals. This is most likely due to the
link between ULXs and star formation (Swartz, Tennant \& Soria 2009), as spiral galaxies typically
display higher star formation rates than elliptical galaxies.

Based on the work of Moretti \etal (2003), we have used the 2XMM sensitivity maps
to estimate the contribution from unidentified contaminants resolved from the cosmic
X-ray background (\eg background AGN) within the catalogue, and find it to be only $\sim$18 per cent.
The fractional contamination is much higher for elliptical galaxies than spiral galaxies,
due to a combination of them being larger in general, and so cover a greater cumulative sky area,
and hosting an intrinsically less numerous ULX population.

\section{High Energy Turnover}
\label{sec_spec_turn}

One of the key observational characteristics of ULX spectra, seen in the highest
quality observations of ULXs to date, is that they display curvature in their
high energy continuum (typically at $\sim$5--6\,\kev); see Stobbart, Roberts
\& Wilms (2006) and Gladstone, Roberts \& Done (2009). Such curvature is not
frequently seen in the spectra of typical XRBs, hence may provide information
on any physical differences between these types of X-ray source. In this section
we attempt to investigate how widespread this spectral downturn is in the
presented ULX population.

A full spectral analysis of each detection included in the catalogue is beyond the
scope of this work, so we turn instead to count rate ratios of the pre-defined
2XMM energy bands, specifically between bands 3, 4 and 5 which cover 1.0--2.0,
2.0--4.5 and 4.5--12.0~\kev\ respectively. We adopt a simple absorbed power-law
model, and generate empirical relations between the count rate ratios of bands 3/4
and bands 4/5 and the photon index $\Gamma$ with PIMMS\footnote{http://heasarc.nasa.gov/docs/software/tools/pimms.html}.
These are used to estimate the band 5 count rate that should be observed assuming
no curvature (constant $\Gamma$) from the band 3/4 ratio, which are in turn compared
to the observed rates. An $n\sigma$ agreement between the two is determined such
that negative values of $n$ imply fewer counts were observed than would be expected,
\ie that the spectrum does display a downturn. We repeat our analysis with estimates
for the average neutral column density of $\langle N_H\rangle = 1$ and $3 \times
10^{21}$\,\atpcm; previous work suggests $\langle N_H\rangle$ should be in this
range (\eg Stobbart \etal 2006; Gladstone \etal 2009; Swartz \etal 2004)

\begin{table}
  \caption{Values for the curvature parameter $C$ (equation \ref{eqn_curv})
for the EPIC PN number distributions of $n$ for the various catalogue subsets
shown in Fig. \ref{fig_turn}. Negative values of $C$ imply a spectral downturn
is observed in the highest 2XMM energy band within the sampled population.}
\begin{center}
\begin{tabular}{c c c}
\hline
\hline
\\[-0.3cm]
& \multicolumn{2}{c}{Curvature Parameter, $C$ ($\times10^{-2}$)} \\
\\[-0.3cm]
\hline
\\[-0.3cm]
$\langle$\nh$\rangle$\tmark[a] & $1\times10^{21}$ & $3\times10^{21}$ \\
\\[-0.3cm]
\hline
\\[-0.3cm]
Whole catalogue & $-17.4\pm2.1$ & $-8.3\pm1.7$ \\
\\[-0.3cm]
\hline
\\[-0.3cm]
Spiral galaxies & $-21.4\pm2.9$ & $-10.1\pm2.3$ \\
\\[-0.3cm]
Elliptical galaxies & $-8.2\pm2.5$ & $-4.1\pm1.7$ \\
\\[-0.3cm]
\hline
\\[-0.3cm]
High quality & $-33.5\pm5.5$ & $-14.3\pm4.3$ \\
\\[-0.3cm]
Medium quality & $-7.5\pm2.2$ & $-2.5\pm1.3$ \\
\\[-0.3cm]
Low quality & $-11.2\pm2.8$ & $-8.1\pm2.3$ \\
\\[-0.3cm]
\hline
\hline
\end{tabular}
\label{tab_curv}
\end{center}
\small $^a$ Column densities are quoted in \atpcm
\end{table}

The \epicpn\ number distributions of $n$ for the whole catalogue and some of its
various subsets are shown in Fig. \ref{fig_turn}. We do not show the distributions
for the \epicmos\ detectors for clarity, but note that they display similar trends.
The top row of the left panel shows the distributions for the whole catalogue,
and it is clear that they are not symmetrically distributed around zero, as would
be expected if no particular spectral curvature was common in the population.
Instead they are skewed and extended towards negative values of $n$, implying
the spectral curvature observed in the best quality ULX datasets may be seen
in a significant fraction of our ULX population. In an attempt to quantify this, we
define the curvature parameter $C$ in equation \ref{eqn_curv} (where $N$ is the
number of detections); $C$ should be consistent with zero for symmetric distributions.
Values are quoted in Table \ref{tab_curv} for all the distributions shown in Fig.
\ref{fig_turn}. The presence of background AGN will have diluted these values, as
they do not typically show curved spectra. In addition to the distributions for the
whole catalogue, the left panel of Fig. \ref{fig_turn} shows a comparison between
sources located in spiral and elliptical galaxies, while the right hand panel shows
the evolution of these distributions with data quality, based on the number of band
5 EPIC counts.

\vspace{0.2cm}
\begin{equation}
C=\frac{N\left ( n\geq 3 \right ) - N\left ( n\leq -3 \right )}{N_{tot}}
\label{eqn_curv}
\end{equation}
\vspace{0.2cm}

It is clear from both the right hand panel in Fig. \ref{fig_turn} and Table
\ref{tab_curv} that the proportion of the population which displays observable
curvature increases with increasing data quality. This is as would be expected
were the curvature intrinsically present in the majority of ULXs; as shown in
Stobbart \etal (2006) it can be subtle to detect even in the
best quality data. Many observations contributing to 2XMM would not have
been long enough to reliably detect continuum curvature at high energies. The best
quality subset of the catalogue, which should provide the most reliable estimate,
suggests this curvature is a common property of ULX spectra. At first, when
comparing the populations from spiral and elliptical galaxies it may seem that
high energy curvature is more common in ULXs located in spiral galaxies. However,
it is likely that $\langle N_H\rangle$ is lower for the elliptical population
than the spiral population. Taking this into consideration, we argue
that high energy curvature is likely to be common to ULXs from both galaxy types.
If so, given that ULX populations in spiral and elliptical galaxies are expected
to be dominated by high- and low-mass binary systems respectively, the likelihood is
that the curvature must be associated with processes in the inner regions
of the X-ray source, rather than any environmental effect.

\section{NGC~4517 ULX1}
\label{sec_dis}

\begin{figure}
\begin{center}
\rotatebox{270}{
{\includegraphics[width=165pt]{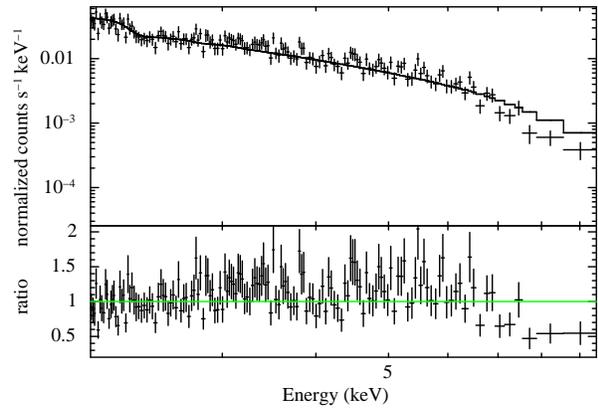}}
}
\end{center}
\caption{The \epicpn\ (black) 2-10\kev\ spectra of \ulxone,
modelled with an absorbed power-law (solid line; see text). This model clearly predicts
an excess of counts over those observed at high energies (above $\sim$6\,\kev). A
data/model ratio plot is also shown in the lower panel.}
\label{fig_ulx1_turn}
\end{figure}

One source that came to light during the production of the catalogue, located in
the edge-on spiral galaxy NGC~4517, is NGC~4517 ULX1. The archived EPIC spectrum
contains $\sim$15500 counts, so the data meet the quality criterion of Stobbart
\etal (2006) and can be considered amongst the best quality ULX data available.
Following their example, we checked the 2.0--10.0~\kev\ spectrum for curvature,
comparing single and broken power-law models. Even allowing for absorption by a
substantial column density of \nh\ $\sim$$8 \times 10^{21}$ \atpcm, as suggested
from inspection of the full 0.3--10.0~\kev\ spectrum, the broken power-law model
is strongly favoured, with a break energy of $\sim$5.5\,\kev, implying the high
energy continuum does indeed display curvature similar to that seen in the other
high quality ULX spectra. This is demonstrated in Fig. \ref{fig_ulx1_turn}.

\begin{figure}
\begin{center}
\rotatebox{0}{
{\includegraphics[width=230pt]{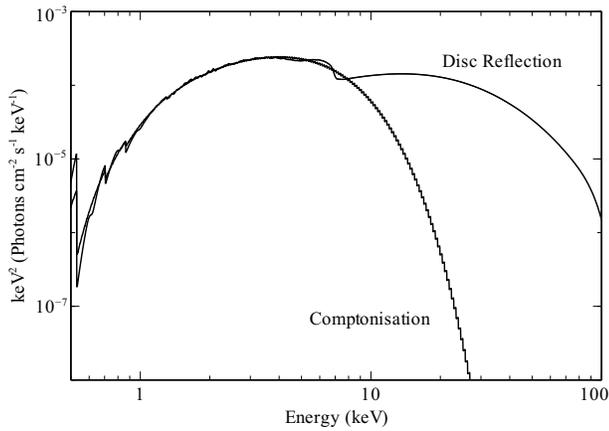}}
}
\end{center}
\caption{A comparison of the Comptonisation and disc reflection interpretations
for ULX spectra; both have been extended beyond the \xmm bandpass up to
100\kev. The models agree well below $\sim$10\,\kev, but there is a clear
discrepancy of many orders of magnitude between them at higher energies, due to
the presence of the Compton hump in reflection spectra.}
\label{fig_comparison}
\end{figure}

A number of explanations of this feature have been proposed, including Comptonisation
of seed photons from the accretion disc in an optically thick corona, which shrouds
the inner regions of the disc (Gladstone \etal 2009) and a combination of relativistically
blurred iron emission and absorption features, originating from the inner regions of
the disc (Caballero-Garc\'ia \& Fabian 2010). Other explanations invoke advection of
radiation in the inner regions of a slim disc (Watarai, Mizuno \& Mineshige 2001),
although these have not been as successful at modelling the broad band X-ray spectra
(see Gladstone \etal 2009). Both the Comptonisation and disc reflection interpretations
are applied to the 0.3--10.0~\kev\ spectrum and found to provide statistically equivalent
representations of the data. We do not provide full details of the spectral modelling here
(see Walton \etal\ 2010, submitted), but merely comment that the results in each case are
similar to those obtained from the application of these models to other ULXs: the electron
temperature of the corona is low and its optical depth is high if Comptonisation is correct,
while the iron abundance of the disc is highly super-solar and the corona very compact (within
a few \rg) if reflection is correct.

A detailed physical consideration of these results is presented in Walton \etal (2010),
but neither model may currently be excluded, so other observational means are required
to distinguish between them. In Fig. \ref{fig_comparison} the two models are
compared after being extended beyond the \xmm\ bandpass to 100~\kev. Above 10~\kev\
the Comptonisation model continues to curve downwards, while the reflection model
turns back up. This is due to the presence of the `Compton hump' in reflection
spectra, a broad emission feature often observed at $\sim$30~\kev, which arises
due to the interplay between photoelectric absorption of low energy photons and
Compton down-scattering of high energy photons within the reflecting medium. The
presence of this feature leads to a large difference in the expected emission
predicted above $\sim$10~\kev\ by the two models. Currently the Comptonisation
model is calculated assuming a purely thermal distribution of electrons, although it
is likely the corona also contains a non-thermal population. Eventually
Comptonisation from this population will arrest the downward curvature,
however this may not become significant until energies  $\gtrsim100$~\kev,
and we still expect there would be large differences between the two model
predictions. Hence, even fairly short observations of ULXs above $\sim$10~\kev\
could be very important in determining the underlying physical processes behind the
observed emission from ULXs.

With current instrumentation such observations are only possible for M~82 X-1
due to the lack of high resolution X-ray imaging spectrometers that operate at
such high energies. The indication from M~82 X-1 is that the spectrum continues
to curve downwards, which favours the Comptonisation interpretation for this
source (Miyawaki \etal 2009). The launch of the hard X-ray imagers aboard
\nustar\footnote{http://by134w.bay134.mail.live.com/default.aspx?wa=wsignin1.0}
and \textit{Astro-H}\footnote{http://astro-h.isas.jaxa.jp/si/index\_e.html} over
the next few years should enable such observations to be carried out for a large
number of ULXs.

\section*{ACKNOWLEDGEMENTS}

DJW is financially supported by STFC, and ACF thanks the
Royal Society. Some of the figures have been produced with
Jeremy Sanders' Veusz\footnote{http://home.gna.org/veusz/} plotting package.

\bibliographystyle{mnras}

\label{lastpage}

\end{document}